\documentclass[aps,prd,twocolumn,floatfix,preprintnumbers,altaffilletter,superscriptaddress]{revtex4-1}
\usepackage{graphicx,url,amssymb,amsmath,rotating,color,units,wasysym,epsfig,multirow,subcaption}
\usepackage{soul,xcolor}
\usepackage[normalem]{ulem}
\usepackage{amsmath}

\newcommand{\red}[1]{{\color{black}#1}}

\begin{document}
\title{Ultra-relativistic astrophysics using multi-messenger observations of double neutron stars with LISA and the SKA}

\author{Eric Thrane}
\email{eric.thrane@monash.edu}
\affiliation{Monash Centre for Astrophysics, School of Physics and Astronomy, Monash University, Clayton, VIC 3800, Australia}
\affiliation{OzGrav: The ARC Centre of Excellence for Gravitational-Wave Discovery, Clayton, VIC 3800, Australia}

\author{Stefan Os\l{}owski}
\email{stefanoslowski@swin.edu.au}
\affiliation{Centre for Astrophysics and Supercomputing, Swinburne University of Technology, Hawthorn VIC 3122, Australia}
\affiliation{OzGrav: The ARC Centre of Excellence for Gravitational-Wave Discovery, Hawthorn, VIC 3122, Australia}

\author{Paul Lasky}
\email{paul.lasky@monash.edu}
\affiliation{Monash Centre for Astrophysics, School of Physics and Astronomy, Monash University, Clayton, VIC 3800, Australia}
\affiliation{OzGrav: The ARC Centre of Excellence for Gravitational-Wave Discovery, Clayton, VIC 3800, Australia}


\begin{abstract}
Recent work highlights that tens of Galactic double neutron stars are likely to be detectable in the millihertz band of the space-based gravitational-wave observatory, LISA.
Kyutoku and Nishino point out that some of these binaries might be detectable as radio pulsars using the Square Kilometer Array (SKA).
We point out that the joint LISA+SKA detection of a $f_\text{gw}\gtrsim\unit[1]{mHz}$ binary, corresponding to a binary period of $\lesssim\unit[400]{s}$, would enable precision measurements of ultra-relativistic phenomena.
We show that, given plausible assumptions, multi-messenger observations of ultra-relativistic binaries can be used to constrain the neutron star equation of state with remarkable fidelity.
It may be possible to measure the mass-radius relation with a precision of \red{$\approx$0.2\%} after $\unit[10]{yr}$ of observations with the SKA. 
Such a measurement would be roughly an order of magnitude more precise than possible with other proposed observations.
We summarize some of the other remarkable science made possible with multi-messenger observations of millihertz binaries, and discuss the prospects for the detection of such objects.
\end{abstract}

\maketitle

\section{Detecting ultra-relativistic Galactic binaries with LISA}
Kyutoku and Nishino~\cite{Kyutoku} recently pointed out that the Laser Interferometer Space Antenna (LISA)~\cite{LISA} is likely to detect ultra-relativistic, Galactic double neutron stars, some of which could be subsequently detected in radio with follow-up from the Square Kilometer Array (SKA)~\cite{SKA}.
These millihertz  binaries have the potential to probe a regime of relativistic astrophysics not accessible with currently known binary systems.
For example, the Double Pulsar (PSR~J0737--3039) has an orbital period of $\unit[2.5]{hr}$ and semi-major axis $a=\unit[6\times10^{-3}]{AU}$~\cite{DoublePulsar}.
In contrast, the double neutron stars observed by LISA with gravitational-wave frequencies $\gtrsim\unit[1]{mHz}$, will have binary periods of $P_B\lesssim\unit[2000]{s}$ and semi-major axes $a\lesssim\unit[1\times10^{-3}]{AU}$~\footnote{We note that LISA is sensitive to gravitational waves at frequencies below $\unit[1]{mHz}$, but we focus on $\gtrsim\unit[1]{mHz}$ since binaries in this band are easier for LISA to detect, and they are more relativistic.}.

The number of double neutron stars emitting gravitational waves above $\unit[1]{mHz}$ can be estimated using the double neutron star merger rate inferred from LIGO/Virgo following the detection of GW170817~\cite{GW170817}.
Following~\cite{Kyutoku,Kyutoku2}, we estimate
\begin{align}
    N_\text{LISA} = & (47-690) 
    \left(\frac{{\cal M}}{\unit[1.2]{M_\odot}}\right)^{-5/3}
    \left(\frac{f_\text{gw}}{\unit[1]{mHz}}\right)^{-8/3} .
\end{align}
Here, ${\cal M}$ is chirp mass, which we take throughout to be $1.2 M_\odot$ (corresponding to an equal mass binary with $1.38 M_\odot$ components).
The range of values (90\% credible interval) comes from uncertainty in the merger rate.
While not all of these binaries will be detectable by LISA, many of them will be.

We adopt the convention that a double neutron star is detectable if it produces a matched-filter signal-to-noise ratio $\rho>7$; see, for example,~\cite{Robson}.
We calculate typical signal-to-noise ratios using~\cite{Seto}
\begin{align}
    \widehat\rho \equiv & \langle\rho^2\rangle^{1/2} \\
    = & \frac{8G^{5/3}T^{1/2}{\cal M}^{5/3}\pi^{2/3}}{5^{1/2}c^4d}
    \left(\frac{f_\text{gw}^{2/3}}{S_n^{1/2}(f_\text{gw})}\right) ,
\end{align}
which is the square root of the signal-to-noise ratio squared, averaged over binary orientation and sky location~\cite{Robson}.
Here, $T$ is the observation time, $G$ is the gravitational constant, $c$ is the speed of light, $d$ is the distance, and $S_n(f_\text{gw})$ is the noise power-spectral density.
We model the LISA noise curve (shown below in Fig.~\ref{fig:asd}) using the $T=\unit[4]{yr}$ prescription from~\cite{Robson}, which includes the effect of foreground from white-dwarf binaries.
Using this expression, one finds that a $\unit[1]{mHz}$ binary can be detected to distances of $d\approx\unit[9]{kpc}$ (beyond the distance to the Galactic Center), while a $\unit[5]{mHz}$ binary can be detected to distances of $d\approx\unit[590]{kpc}$, 75\% the distance to Andromeda.

This result is roughly consistent with other estimates of the number of double neutron stars detectable with LISA.
Kremer et al.~\cite{NU} examined the population of LISA-band binaries in the globular clusters of the Milky Way.
They estimate $22$ globular-cluster double neutron stars will radiate in the LISA band ($\approx\unit[10^{-2}-100]{mHz}$).
Of these, two double neutron stars are likely to be detected above the LISA noise floor.
Since many binaries in globular clusters form dynamically, many of these systems have significant eccentricity.
The number of millihertz binaries in globular clusters is likely to be small compared to the number of millihertz binaries in the field since the prevalence of millihertz binaries is directly related to the double neutron star merger rate, and $N$-body studies predict that globular clusters are relatively inefficient at merging double neutron stars; see, e.g.,~\cite{Belczynski}.
See also work by Seto~\cite{Seto} and~Lau et al.~\cite{Lau}, who considered the population of LISA-band double neutron stars in the Local Group.

Our best guess for the Galactic rate of double neutron star mergers is $\unit[1.5\times10^{-4}]{MWEG^{-1} yr^{-1}}$~\cite{GW170817}, which implies a typical time between mergers in the Milky Way of $\sim\unit[6700]{yr}$.
(Here, MWEG stands for ``Milky Way Equivalent Galaxy.'')
Therefore, our best guess for the shortest time to merge for a Galactic double neutron star binary is half that: $\unit[3300]{yr}$.
Below, we focus our attention on binaries that are, in principle, observable as pulsars.
Here, we assume that $\approx10\%$ of the recycled neutron stars in double neutron star systems are detectable as pulsars due to beaming effects.
Thus, our best guess for the shortest time to merge for a Galactic double neutron star binary {\em with a potentially observable radio pulsar} is \red{$\unit[33]{kyr}$}.

Let us imagine that such a double neutron star binary exists in the Milky Way and give it a fictitious name: PSR~J1234--5678 or ``J1234'' for short.
We proceed to investigate the properties of J1234, and then see how our results would change assuming different times to merge.
For reasons that will become clear momentarily, we are especially interested in millihertz binaries with non-negligible eccentricity.
Therefore, let us further suppose that J1234 was born recently through unstable ``case BB'' mass transfer.
Such systems have been hypothesized as possible progenitors for binary neutron star mergers~\cite{Ivanova,Belczynski2,VignaGomez} as well as sources for $r$-process enrichment in ultra-faint dwarf galaxies~\cite{Safarzadeh}.
In this scenario, a neutron star - helium star binary undergoes unstable mass transfer, leading to a common envelope event, and eventually---in some cases---a neutron star - helium core binary with an orbital period of ${\cal O}(\unit[1000]{s})$.
Since the binary is so tight at this stage of its evolution, it is likely to survive when the helium core undergoes a supernova, leading to a double neutron star binary likely to merge in $\lesssim\unit[10]{Myr}$.
A binary with such a short life time can retain significant eccentricity as it passes through the LISA band.
For illustrative purposes, we assume that J1234 was born with an eccentricity of \red{$e_0=0.75$} (typical of population synthesis studies) and a period of \red{$P_b^0=\unit[12]{ks}$} giving it a lifetime of $\unit[10]{Myr}$~\cite{Peters64}.

We evolve J1234 forward in time using the standard prescription from~\cite{Peters64}, so that it is \red{$\unit[33]{kyr}$} from merger.
At this point in its evolution, the orbital period of J1234 is \red{$P_b = \unit[490]{s}$} and the eccentricity is \red{$e=0.11$}, which is comparable to the Double Pulsar.
Approximating the orbit as quasi-circular, the gravitational-wave frequency is \red{$\unit[4.1]{mHz}$}.
In the remainder of this note, we consider some of the science possible by multi-messenger study of J1234.
(In our closing remarks, we discuss how the results would change given different assumptions about J1234.)

In Fig.~\ref{fig:asd}, we show the effective gravitational-wave strain from J1234 relative to the LISA noise curve assuming a typical distance of $d=\unit[10]{kpc}$ and an observation time of $T=\unit[4]{yr}$
\begin{align}
    h_\text{eff}^2 \equiv \frac{32}{5}f_\text{gw}^2TS_h(f),
\end{align}
where
\begin{align}
    S_h(f) = \frac{{\cal A}^2(f)}{2T} ,
\end{align}
and
\begin{align}
    {\cal A}^2(f) = \frac{5}{24}\frac{(G{\cal M}/c^3)^{5/3}f_\text{gw}^{-7/3}}{\pi^{4/3}d^2/c^2} .
\end{align}
As noted above, the noise curve is calculated using the $T=\unit[4]{yr}$ prescription from~\cite{Robson}.
We include instrumental noise (red) as well as confusion noise from the white dwarf background (blue).
The effective strain from J1234 is indicated with the green star.
Approximating the orbit as quasi-circular, the gravitational waves from J1234 would produce a matched filter signal-to-noise ratio of \red{$\rho=360$}.
(This implies a $\rho=7$ detection distance of \red{$\unit[510]{kpc}$}, about 67\% the distance to Andromeda).
The dotted green line shows the evolution of the binary from $\unit[1.06]{mHz}$ (\red{$\unit[800]{kyr}$} away from merger), when the binary would have first become detectable to LISA with $\rho\gtrsim7$.

LISA is capable of precise localisation because of the large baseline created from its orbit around the Sun.
Monochromatic sources like J1234 can be localised to within a solid angle of~\cite{Kyutoku}
\begin{align}
    \Delta\Omega = \unit[0.036]{deg^2}
    \left(\frac{\rho}{200}\right)^{-2}
    \left(\frac{f_\text{gw}}{\unit[4]{mHz}}\right)^{-2} ,
\end{align}
which works out to \red{$\unit[0.011]{deg^2}$} for J1234, well within a single SKA beam.
The distance is measured to a fractional uncertainty of of~\cite{Kyutoku}
\begin{align}
    \frac{\Delta d}{d} = 0.01
    \left(\frac{\rho}{200}\right)^{-1} ,
\end{align}
which implies that the distance to J1234 will be known from LISA measurements to \red{0.6\%}.
Knowing the distance so precisely will be useful in order to control syetmatic error from proper motion.

\begin{figure}
\centering
\includegraphics[width=0.49\textwidth]{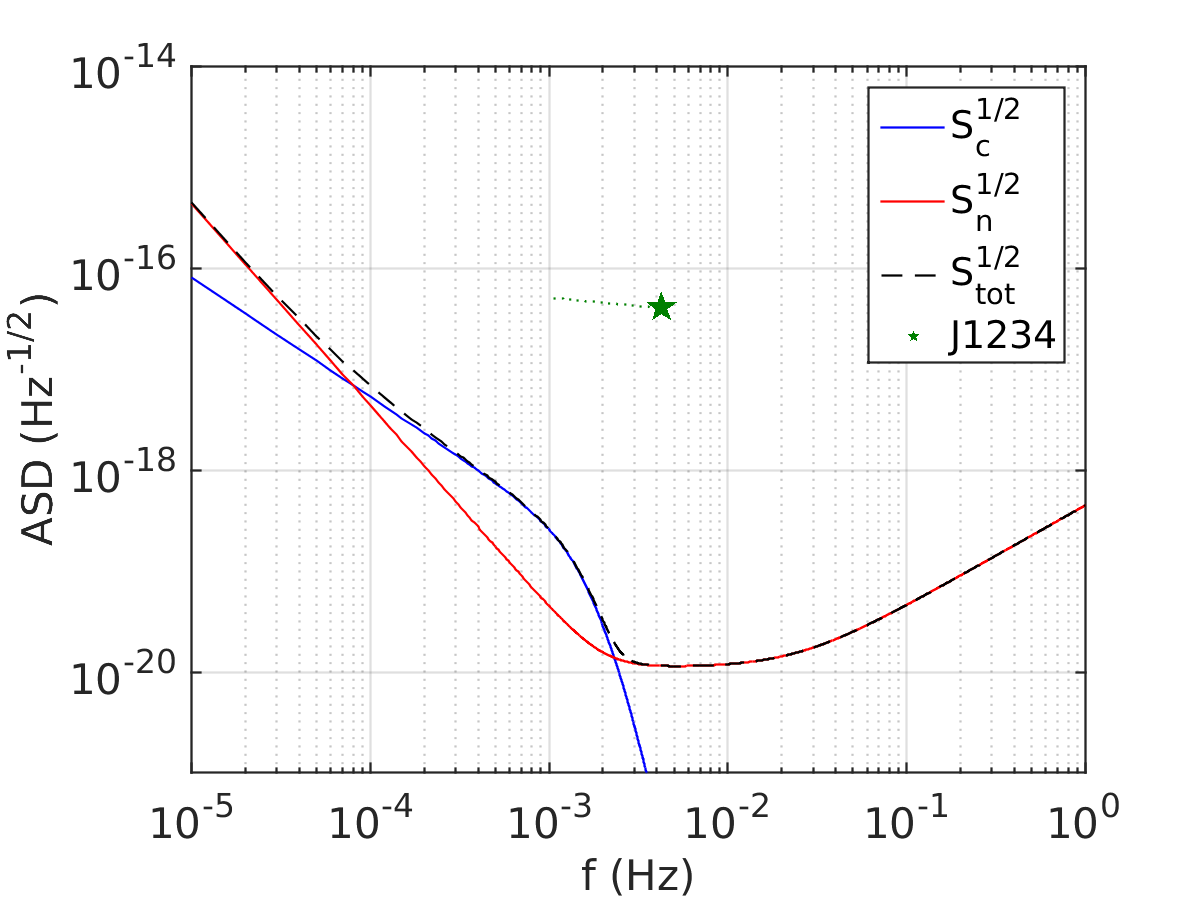}
\caption{
The gravitational-wave effective strain from J1234 relative to the LISA amplitude spectral density noise curve.
The confusion noise from white dwarf binaries is shown in blue.
The instrumental noise is shown in red.
The total noise is black.
The effective strain from J1234 is shown as a green star assuming a distance of $d=\unit[10]{kpc}$.
The dotted green line shows the evolution of J1234 from $\unit[1]{mHz}$ when it would have first been detectable by LISA.
}
\label{fig:asd}
\end{figure}

The early phase of SKA is expected to quadruple the population of known pulsars in the Galaxy, and the full SKA will detect all pulsars in our Galaxy beaming towards Earth, thus increasing the sample of known double neutron stars by a factor of 5-10 \citep{Keane2018,Levin2018}.
While population synthesis studies often adopt the work of Tauris and Manchester \citep{TM98} to estimate the beaming fraction, this is an optimistic assumption for double neutron star systems as the relation between the period and beaming fraction is a best fit for pulsars with longer periods.
Double neutron star systems are likely to have tighter beams.
For example, the recycled star in the Double Pulsar system has a beaming fraction of about 10\%, and accordingly only one in ten double neutron star systems might be seen by the SKA.
A further complication of finding millihertz binaries is that they are highly accelerated.
While significant progress has been made in this area \citep[e.g.,][]{Cameron18}, this still remains a computationally challenging aspect of pulsar searching.
However, if J1234 was discovered first by LISA, a targeted search would largely alleviate this problem.
In the discussion that follows we assume that J1234 is detectable by the SKA.


\section{Lense-Thirring Precession}
Let us suppose that one of the neutron stars in J1234 is rapidly rotating.
This rotation will give rise to Lense-Thirring precession.
Lense-Thirring precession is one of four sources of periastron advance contributing to the total periastron advance~\cite{Bagchi} of a pulsar $a$ with companion $b$
\begin{align}
    \dot\omega_\text{tot}^a =
    \dot\omega_\text{PN}^a +
    \dot\omega_\text{secular} +
    \dot\omega_\text{LT}^a + 
    \dot\omega_\text{LT}^b .
\end{align}
Here, $\dot\omega_\text{PN}^a$ is the post-Newtonian correction at $a$ due to the spacetime curvature from both $a$ and $b$; $\dot\omega_\text{secular}$ is secular variation due to the change in the apparent orientation of the binary with respect to the line of sight because of the proper motion of the binary barycenter; $\dot\omega_\text{LT}^a$ is the Lense-Thirring precession from $a$; and $\dot\omega_\text{LT}^b$ is the Lense-Thirring precession from $b$.

We make a number of simplifying assumptions.
We assume that: (i) $b$ is slowly rotating so that $\dot\omega_\text{LT}^b=0$; (ii) the neutron stars have essential equal masses; (iii) the binary is viewed edge-on with inclination angle $\iota=90^\circ$; (iv) the spin vector of pulsar $a$ is  aligned to the total angular momentum.
Given these assumptions, and following the calculation from~\cite{Bagchi}, the advance of periastron from Lense-Thirring precession is~
\begin{align}
    \dot\omega_\text{LT}^a = & 14\pi^3 \frac{I_a}{P_b^2 P_a M c^2} .
\end{align}
Here, $I_a$ is the moment of inertia for pulsar $a$, $P_b$ is the binary period, and $P_a$ is the spin period of $a$.
Plugging in plausible values for the parameters of J1234, we obtain
\begin{align}
    \dot\omega_\text{LT}^\text{J1234} = & 
    \red{\unit[4.1\times10^{-2}]{deg\,yr^{-1}}} \Bigg[
    \left(\frac{I_a}{\unit[1.26\times10^{45}]{g\,cm^2}}\right) \nonumber\\
    & \left(\frac{\red{\unit[490]{s}}}{P_b}\right)^2 
    \left(\frac{\unit[20]{ms}}{P_a}\right) 
    \left(\frac{\unit[2.76]{M_\odot}}{M}\right) \Bigg] .
\end{align}

Previous work has studied how the SKA will be able to measuring Lense-Thirring precession in the Double Pulsar~\cite{Kehl}.
According to these authors, the SKA will achieve a Double Pulsar precession sensitivity of  $\dot\omega=\unit[10^{-4}]{deg\,yr^{-1}}$ with after four years of SKA1 and $\dot\omega=\unit[2\times10^{-5}]{deg\,yr^{-1}}$ after ten years of SKA (4 years of SKA1 + an additional 6 years of SKA at design sensitivity).
While the magnitude of Lense-Thirring precession is greater for short-period binaries, scaling like $\dot\omega_\text{LT}\propto P_b^{-2}$, the pulsar-timing signal scales like the semi-major axis $\tau\propto a\sin\iota$.
Taking into account these two effects, and applying Kepler's third law,the signal-to-noise ratio for Lense-Thirring precession scales as
\begin{align}
    \text{SNR} \propto & P_b^{-4/3} \sin\iota .
\end{align}
Naively applying this scaling law---we discuss caveats momentarily---we estimate that the precession of J1234 can be measured with signal-to-noise ratio of SNR=\red{$60$} after four years of operation and SNR=\red{$300$} after 10 years of operation.

The precision of timing experiments is predominantly determined by the sensitivity of the telescope relative to the flux of the pulsar and the spin period of the pulsar.
The Double Pulsar spin period is quite typical for a double neutron star system \citep{Oslowski2011}. The measurement of binary period derivative is largely limited due to various effects affecting the observed magnitude, such as the Shklovskii effect and the uncertainty of the Galactic potential.
However, the combination of a good estimate of distance to J1234 combined with a model of the Galactic potential to large distances from Earth provided by Gaia and other surveys \citep{Gaia2016} will mitigate these potential problems.

The main special property of the Double Pulsar is its near edge-on alignment and the sweeping of the pulsar A's beam across the magnetosphere of pulsar B, enabling exciting physical experiments \citep{Breton2008}. 
However, neither of these two conditions are prerequisite for measuring precession.
Another property of the Double Pulsar, which may not be unique, but which is relevant here, is that J0737-3039A is an aligned rotator and, not affected by geodetic precession.
If that was not the case in J1234, the precession period would be about 150 times shorter than for the Double Pulsar \citep{Stairs2003}, which means it would precess about twice a year.

In this case, timing measurements of J1234 could be adversely affected by reduced cadence.
However, given the expected length of the observing campaign, and a good coverage of the orbit thanks to the short orbital period, the timing would still provide a good measurement if a phase-connected solution can be found between epoch in which the pulsar is beaming towards Earth.
Moreover, precession can increase the fraction of binaries beaming towards Earth while helping to constrain other binary parameters of the system.
Here, we assume that J1234 is either an aligned rotator, or that the impact of precession on the timing measurement of required parameters is not deleterious.
Therefore, our naive scaling argument provides a reasonable approximation for our ability to measure Lense-Thirring precession in J1234.


\section{Equation of State}
Previous work has noted that Lense-Thirring precession in the Double Pulsar can be used to constrain the neutron star equation of state~\cite{Damour,Kehl,Bagchi}.
The neutron-star moment of inertia depends on the mass, the radius, and the equation of state~\cite{Ravenhall}.
Since the mass of J1234 and its companion are precisely determined using pulsar timing, one can compare the measured moment of inertia with the moment of inertia implied by $M$ and $R(M)$.
The uncertainty in $R(M)$ can be approximated as
\begin{align}
    \frac{\delta R(M)}{R(M)} \approx \frac{1}{2} \frac{\delta I}{I} .
\end{align}
Thus, an SNR=\red{300} measurement of $\dot\omega$ will constrain $R(M)$ for one value of $M$ to a precision of \red{$\approx$0.2\%}.
We note that there are unlikely to be many such multi-messenger measurements of ultra-relativistic double neutron stars because there are a limited number in the LISA-band.
Thus, other measurements of the neutron star equation of state, for example, from LIGO~\cite{GW170817_eos} and/or NICER~\cite{nicer}, will be needed to supplement constraints on $R(M)$ from ultra-relativistic binaries.

In order to investigate how this result is affected by uncertainty in the double neutron star merger rate, we repeat the calculation above using the extreme values of the 90\% rate credible interval.
A minimum (maximum) rate of $\unit[320]{Gpc^{-3}yr^{-1}}$ ($\unit[4740]{Gpc^{-3}yr^{-1}}$) implies a J1234 merger time of \red{$\unit[3.2\times10^5]{yr}$ ($\unit[2.2\times10^4]{yr}$)}~\footnote{Note, this time to merger includes the aforementioned factor of ten penalty, which takes into account the fact that not all double neutron stars will are likely to include a visible pulsar.} 
These times to merger imply gravitational-wave frequencies of \red{$f_\text{gw}=\unit[1.6]{mHz}$ ($f_\text{gw}=\unit[4.7]{mHz}$)}.
Even at a reduced frequency of \red{$f_\text{gw}^\text{min}=\unit[1.6]{mHz}$}, J1234 could be observed out to a distance of \red{$d=\unit[42]{kpc}$} (most of the Milky Way).
The eccentricity in the LISA band would be \red{$e=0.25$ ($e=0.09$)}.
%
The resulting Lense-Thirring signal would change to \red{$\dot\omega_\text{LT}=\unit[0.03]{deg\,yr^{-1}}$ ($\unit[0.28]{deg\,yr^{-1}}$)}, which corresponds to a signal-to-noise ratio of \red{SNR=46 (SNR=400)} after ten years of SKA.
Thus, taking into account Poisson uncertainty from the double neutron star merger rate, we calculate that the neutron star $R(M)$ can be constrained with J1234 at the level of approximately \red{$0.1\%-1\%$} (90\% credible interval) using ten years of SKA.


\section{Discussion}
Double neutron stars represent an exciting multi-messenger source for LISA and the SKA.
LISA is adept at detecting millihertz double neutron stars that would be challenging to detect in the radio owing to their large accelerations.
Once the binary has been identified by LISA, the SKA can (1) separate out double neutron stars from binaries with one or more white dwarf and (2) measure the binary evolution with a sensitivity that is many orders of magnitude better than LISA.
Such ultra-relativistic binaries provide a unique laboratory for a variety of tests, including tests of general relativity and the composition of neutron stars.

While neutron stars have received considerable attention in the SKA literature, they are to some extent overshadowed by black hole science in the LISA literature.
In light of the science made possible by multi-messenger observations by LISA and the SKA, it seems that there are exciting new lines of inquiry worth pursuing in the future.
We highlight a couple here.
First, using the SKA and LISA, it may be possible to study the population properties of double neutron stars in the Milky Way, studying the population changes with binary period.
Second, LISA is capable of detecting the most ultra-relativistic double neutron stars like J1234 ($\gtrsim\unit[10]{mHz}$) in nearby galaxies.
Using ephemerides provided by LISA, it may be possible to observe these pulsars with the SKA as well; see, for example,~\cite{Kramer}.

We thank Ilya Mandel and Simon Stevenson for helpful discussions on millihertz binaries.
This work is supported through Australian Research Council (ARC) Future Fellowships FT150100281, FT160100112, FL150100148, Centre of Excellence CE170100004, and Discovery Project DP180103155.

\bibliography{bibliography}

\end{document}